%% file: OnSecureDistributedDataStorageUnderRepairDynamics.tex
\documentclass[10pt,conference]{IEEEtran}

\usepackage{times}
\usepackage{epsfig}
\usepackage{amsmath}
\usepackage{amsfonts}
\usepackage{graphicx}
\usepackage{amssymb}
\usepackage{amstext}
\usepackage{latexsym}
\usepackage{color}
\usepackage{ifthen}
\usepackage{multirow}
\usepackage{verbatim}

\newcommand{\be}{\begin{equation}}
\newcommand{\ee}{\end{equation}}
\newcommand{\bear}{\begin{eqnarray}}
\newcommand{\eear}{\end{eqnarray}}
\newcommand{\bears}{\begin{eqnarray*}}
\newcommand{\eears}{\end{eqnarray*}}
\newcommand{\bi}{\begin{itemize}}
\newcommand{\ei}{\end{itemize}}
\newcommand{\ben}{\begin{enumerate}}
\newcommand{\een}{\end{enumerate}}

\newcommand{\vsp}{\vspace*{-0.25in}}
\newtheorem{theorem}{Theorem}

\renewcommand{\baselinestretch}{0.95}
\begin{document}

\title{On Secure Distributed Data Storage Under Repair
Dynamics}
%

\author{ Sameer Pawar, Salim El Rouayheb, Kannan Ramchandran\\
 University of California, Berkeley\\
 Emails: \texttt{\{spawar,salim,kannanr\}@eecs.berkeley.edu}.}

\long\def\symbolfootnote[#1]#2{\begingroup%
\def\thefootnote{\fnsymbol{footnote}}\footnote[#1]{#2}\endgroup}

\maketitle
\begin{abstract}
We address the problem of securing distributed storage systems
against passive eavesdroppers that can observe a limited number of
storage nodes. An important aspect of these systems is node failures
over time, which demand a repair mechanism aimed at maintaining a
targeted high level of system reliability. If an eavesdropper observes a node
that is added to the system to replace a failed node, it will have
access to all the data downloaded during repair, which can
potentially compromise the entire information in the system. We are
interested in determining the \emph{secrecy capacity} of distributed
storage systems under repair dynamics, i.e., the maximum amount of
data that can be securely stored and made available to a legitimate
user without revealing any information to any eavesdropper. We derive
a general upper bound on the secrecy capacity and show that this
bound is tight for the \emph{bandwidth-limited regime} which is
of importance in scenarios such as peer-to-peer distributed
storage systems. We also provide a simple explicit code construction
that achieves the capacity for this regime.
\end{abstract}

\symbolfootnote[0]{This research was funded in part by an AFOSR grant (FA9550-09-1-0120), a DTRA grant (HDTRA1-09-1-0032),
and an NSF grant  (CCF-0830788).}

\section{Introduction}\label{sec:Intro}
Data storage devices have evolved significantly since the days of
punched cards. Nevertheless, storage devices, such as hard disks or
flash drives, are still bound to fail after long periods of usage,
risking the loss of valuable data. To solve this problem and to
increase the reliability of the stored data, multiple storage nodes
can be networked together to redundantly store the data, thus forming
a distributed data storage system. Applications of such systems are
innumerable and include large data centers and peer-to-peer storage
systems, such as OceanStore \cite{Ocean}, that use a large number of
nodes spread widely across the Internet to store files.

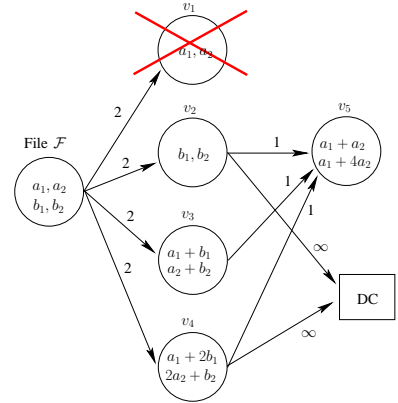
\begin{figure}[t]
\begin{center}
\resizebox{2in}{!} {\input{dss_example.pstex_t}} \caption{An example
of a distributed data storage system under repair. A file
$\mathcal{F}$ of 4 symbols $(a_1,a_2,b_1,b_2)\in \mathbb{F}_5^4$ is
stored on four nodes using an MDS code. Node $v_1$ fails and is
replaced by a new node $v_5$ that downloads $(b_1+b_2),
(a_1+a_2+b_1+b_2)$ and $(a_1+4a_2+2b_1+2b_2)$ from $v_2$, $v_3$, and
$v_4$ respectively to compute and store $(a_1 + a_2, a_1+4a_2)$.
Nodes $v_2,\dots,v_5$ form a new MDS code. The edges in the graph are
labeled by their capacities. The figure also depicts a data
collector connecting to nodes $v_2$ and $v_4$ to recover the stored
file.}\label{fig:distributed_system}\vsp
\end{center}
\end{figure}

Codes for protecting data from erasures have been well studied in
classical channel coding theory, and can be used here to increase the
reliability of distributed storage systems.
Fig.~\ref{fig:distributed_system} illustrates an example where
a maximal distance separable (MDS) code is used to store a file
$\mathcal{F}$ of 4 symbols, $(a_1,a_2,b_1,b_2)\in \mathbb{F}_5^4$,
distributively on $4$ nodes, $v_1,\dots,v_4$, each of capacity $2$
symbols. The MDS code implemented here ensures that any user, also
called data collector, connecting to any $2$ storage nodes can obtain
the whole file $\mathcal{F}$. However, what distinguishes the
scenario here from the erasure channel counterpart is that when a
storage node fails, it needs to be repaired or replaced by a new node
in order to maintain a desired level of system reliability. A
straightforward repair mechanism would be to add a new replacement
node of capacity 2, and make it act as a data collector by
connecting to $2$ surviving nodes. The new node can then download the
whole file (4 symbols) to construct the lost part of the data and
store it. Another repair scheme that consumes less bandwidth is
depicted in Fig.~\ref{fig:distributed_system} where node $v_1$ fails
and is replaced by node $v_5$. When node $v_5$ connect to 3 nodes
instead of 2, it is possible to decrease the total repair bandwidth
from 4 to 3 symbols. Note that $v_5$ does not need to store the
exact data that was on $v_1$; the only required property is that the
data stored on all the active nodes $v_2,v_3,v_4$ and $v_5$ form an
MDS code.

The above important observations were the basis of the original work
of \cite{DGWWR07} where the authors showed that there exists a
fundamental tradeoff between the storage capacity of each node and
the repair bandwidth. They also introduced and constructed
``regenerating codes'' as a new class of codes that generalize
classical erasure codes and permit the operation of a distributed
storage system at any point on the tradeoff curve.
When a distributed data storage system is formed using nodes widely
spread across the Internet, e.g., Internet based peer-to-peer
systems, individual nodes may not be secure and can become
susceptible to eavesdropping. This paper focuses on such scenarios
where an eavesdropper can gain access to a certain number of the
storage nodes. The compromised distributed storage system is always
assumed to be dynamic with nodes continually failing and being
repaired. Thus, the compromised nodes can belong to the original set
of storage nodes that the system starts with, or even include some of
the replacement nodes added to the system to repair it from failures.
Under this setting, we are interested in determining how much data
can still be stored in the system without revealing any information
to any of the eavesdroppers.

To answer this question, we follow the approach of \cite{DGWWR07} and
model the distributed storage system as a multicast network that uses
network coding. Under this model, the eavesdropper is an intruder
that can access a fixed number of the network nodes of her choice.
This eavesdropper model is natural for distributed storage systems
and comes in contrast with the wiretapper model studied in the
network coding literature \cite{CY02, RS07, SK08} where the intruder
can observe network edges, instead of nodes. We derive a general upper
bound on the secrecy capacity as a function of the node storage
capacity and the repair bandwidth. Motivated by system
considerations, we define an important operating regime, that we call
the \emph{bandwidth-limited regime}, where the repair bandwidth
is constrained not to exceed a given upper bound, while no limitation is imposed on
the storage capacity of the nodes. For this important operating
regime, we show that our upper bound is tight and present capacity-achieving codes.

This paper is organized as follows. In Section~\ref{sec:model} we
describe the system and security model. We define the problem and
give a summary of our results in Section~\ref{sec:problem}. In
Section~\ref{sec:special} we illustrate two special cases of
distributed storage systems that are instructive in understanding the
general problem. In Section~\ref{sec:UB_capacity}, we derive an upper
bound on the secrecy capacity, and in Section
\ref{sec:achievability}, we present a scheme that achieves this upper
bound for the case of bandwidth-limited regime. We conclude in
Section~\ref{sec:con}.

\section{Model}\label{sec:model}
\subsection{Distributed storage system}
A distributed storage system (DSS) is a collection of storage nodes
that includes a source node $s$, that has an incompressible data file
$\mathcal{F}$ of $R$ symbols, or units, each belonging to a finite
field $\mathbb{F}$. The source node is connected to $n$ storage nodes
$v_1,\dots,v_n,$ each with a storage capacity of $\alpha$ symbols,
which may be utilized to save coded parts of the file $\mathcal{F}$.
The storage nodes are individually unreliable and may fail over time.
To guarantee a certain desired level of reliability, we assume that
the DSS is required to always have $n$ active, i.e., non-failed,
storage nodes that are in service. Therefore, when a storage node
fails, it is immediately replaced by a new node with same storage
capacity $\alpha$. The DSS should be designed in a way to allow any
legitimate user, that we also call data collector, that connects to
any $k$ out of the $n$ active storage nodes available at any given
time, to be able to reconstruct the original file $\mathcal{F}$. We
term this condition as the ``\emph{reconstruction property}'' of
distributed storage systems.

We assume that nodes fail one at a time, and we denote by $v_{n+i}$
the new replacement node added to the system to repair the $i$-th
failure. The new replacement node connects then to some $d$ nodes,
chosen randomly, out of the remaining active $n-1$ nodes and
downloads $\gamma$ units from them in total, which corresponds to the
{\em repair bandwidth} of the system. The repair degree $d$ is a
system parameter satisfying $k\leq d\leq n-1$. In this work, we focus
on the case of symmetrical repair where the new node downloads equal
amount of data, say $\beta$ units, from each of the $d$ nodes it
connects to, i.e., $\gamma = d\beta$. The process of replenishing
redundancy to maintain the reliability of a DSS is referred to as the
\emph{``regeneration"} or \emph{``repair"} process. Note that a new
replacement node may download more data than what it actually stores.
Moreover, the stored data can possibly be different than the one
that was stored on the failed node, as long as the ``reconstruction
property" of the DSS is retained. A distributed storage system
$\mathcal{D}$ is thus characterized as $\mathcal{D}(n,k)$. For
instance, the DSS depicted in Fig.~\ref{fig:distributed_system}
corresponds to $\mathcal{D}(4,2)$ which is operating at $(\alpha,
\gamma)=(2,3)$.

\subsection{Flow Graph Representation}
\begin{figure}[t]
\begin{center}
\resizebox{2.2in}{!} {\input{flowgraph_ex.pstex_t}} \caption{The flow
graph model of the DSS ${\cal D}(4,2)$, with $d=3$, of
Fig.~\ref{fig:distributed_system} when node $v_1$ fails and is
replaced by node $v_5$. Each storage node $v_i$ is represented by two
nodes $x_{in}^i$ and $x_{out}^i$ connected by an edge
$(x_{in}^i,x_{out}^i)$ of capacity $\alpha$ representing the node
storage constraint. A data collector DC connecting to nodes $v_2$
and $v_4$ is also depicted.}\label{fig:FlowGraph}\vsp
\end{center}
\end{figure}
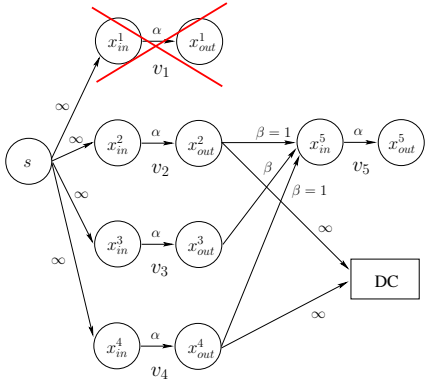

We adopt the flow graph model introduced in \cite{DGWWR07} which we
describe here for completeness. In this model, the distributed
storage system is represented by an information flow graph
$\mathcal{G}$. The graph $\mathcal{G}$ is a directed acyclic graph
with capacity constrained edges that consists of three kinds of
nodes: a single source node $s$, input storage nodes $x^i_{in}$ and
output storage nodes $x^i_{out}$ and data collectors DC$_j$ for $i,j
\in \{1,2,\dots\}$. The source node $s$ has an information $S$ of
which a specific realization is the file $\mathcal{F}$. Each
storage node $v_i$ in the DSS is represented by two nodes $x^i_{in}$
and $x^i_{out}$ joined by a directed edge of capacity $\alpha$ (see
Fig.~\ref{fig:FlowGraph}), to account for the node storage constraint.

The repair process is initiated every time a failure occurs. As a
result, the DSS, and consequently the flow graph, are dynamic and
evolve with time. At any given time, each node in the graph is
either active or inactive depending on whether it has failed or not. The graph $\mathcal{G}$ starts with only the source node $s$
being active and connected to the storage input nodes
$x^1_{in},\dots, x^n_{in}$ by outgoing edges of infinite capacity.
From this point onwards, the source node $s$ becomes and remains
inactive and the $n$ input and output storage nodes become active.
When a node $v_i$ fails in a DSS, the corresponding nodes $x_{in}^i$
and $x_{out}^i$ become inactive in $\mathcal{G}$. If a replacement
node $v_j$ joins the DSS in the process of repairing a failure and
connects to $d$ active nodes $v_{i_1},\dots,v_{i_d}$, the
corresponding nodes $x_{in}^j$ and $x_{out}^j$, with the edge
$(x_{in}^j,x_{out}^j)$, are added to the flow graph $\mathcal{G}$,
and node $x_{in}^j$ is connected to the nodes
$x_{out}^{i_1},\dots,x_{out}^{i_d}$ by incoming edges of capacity
$\beta$ each. A data collector is represented by a node connected
to $k$ active storage output nodes through infinite capacity links
enabling it to reconstruct the file $\mathcal{F}$. The graph
$\mathcal{G}$ constitutes a multicast network with the data
collectors as destinations. An underlying assumption here is that the
flow graph corresponding to a distributed storage system depends on
the sequence of failed nodes. As an example, we depict in
Fig.~\ref{fig:FlowGraph} the flow graph corresponding to the DSS
$\mathcal{D}(4,2)$ of Fig.~\ref{fig:distributed_system}, when node
$v_1$ fails.
\subsection{Eavesdropper Model}
We assume the presence of an intruder ``Eve" in the DSS, who can
observe up to $\ell$, $ \ell < k,$ nodes of her choice among all the
storage nodes, $v_1,v_2,\dots,$ possibly at different time instances
as the system evolves. In the flow graph model, Eve is an
eavesdropper who can access a fixed number $\ell$ of nodes chosen
from the storage input nodes $x_{in}^1,x_{in}^2,\dots$. Notice that
while a data collector observes output storage nodes, i.e., the data
stored on the nodes it connects to, Eve, has access to input storage
nodes, and thus can observe, in addition to the stored data, all
incoming messages to these nodes. We also assume that Eve has
complete knowledge of the storage and repair schemes implemented in
the DSS. Thus, she can choose some of the $\ell$ nodes to be among
the initial $n$ storage nodes, or, if she deems it more profitable,
she can choose to wait for failures and eavesdrop on a replacement
node by observing its downloaded data. Eve is assumed to be passive,
and only observes the data without modifying it.

\section{Problem Statement and Results}\label{sec:problem}

\subsection{Secrecy Capacity}
Let $S$ be a random vector uniformly distributed over
$\mathbb{F}_q^R$, representing the incompressible data file
at the source node with $H(S)= R$. Let
$V_{in}:=\{x_{in}^1,x_{in}^2,\dots\}$ and
$V_{out}:=\{x_{out}^1,x_{out}^2,\dots\}$ be the sets of input and
output storage nodes in ${\cal G}$ respectively. For a storage node
$v_i$, let $D_i$ and $C_i$ be the random variables representing its
downloaded messages and stored content respectively. Thus, $C_i$,
represents the data that can be downloaded by a data collector when
contacting node $v_i$, while $D_i$, with $H(D_i)\leq \gamma$,
represents the total data revealed to Eve when she accesses node
$v_i$. The stored data $C_i$ is a function of the downloaded data
$D_i$.

Let $V_{out}^{a}$ be the collection of all subsets of $V_{out}$ of
cardinality $k$ consisting of nodes that are simultaneously active at
some instant in time. For any subset $B$ of $V_{out}$, define
$C_B:=\{C_i:x_{out}^i\in B\}$. Similarly, for any subset $E$ of
$V_{in}$, define $D_E:=\{D_i:x_{in}^i\in E\}$. The reconstruction
property, then, can be written as
\begin{eqnarray}\label{eq:reconstruction_condition}
H(S|C_{B}) &=& 0\quad \forall B\in V_{out}^{a},
\end{eqnarray} and the perfect secrecy condition implies
\begin{eqnarray}\label{eq:secrecy_condition}
H(S|D_{E}) &=& H(S), \forall E\subset V_{in} \text{ and } |E|\leq
\ell.
\end{eqnarray}
Given a DSS $\mathcal{D}(n,k)$ with $\ell$ compromised nodes, its
secrecy capacity, denoted by $C_s(\alpha,\gamma$), is then defined to
be the maximum amount of data that can be stored in this system such
that the reconstruction property and the perfect secrecy condition
are simultaneously satisfied for all possible data collectors and
eavesdroppers i.e.,
\begin{eqnarray}\label{eq:secrecy_capacity}
C_s (\alpha,\gamma):= \sup_{{\small \left.
                 \begin{array}{cc}
                  H(S|{C}_{B}) = 0 & \forall B\\
                  H(S|{D}_{E}) = H(S) & \forall E\\
                 \end{array}
                \right.}} H(S)
                \end{eqnarray}
where $B\in V_{out}^a$, $E\subset V_{in}$ and $|E| \leq \ell.$

\subsection{Results}
First, we give the following general upper bound on the secrecy
capacity of a DSS:
\begin{theorem}\label{thm:converse}{[Upper Bound]} For a distributed data storage
system $\mathcal{D}(n,k)$, with a repair degree $d$, and $\ell<k$
compromised nodes, the secrecy capacity is upper bounded as
\begin{equation}\label{eq:UB}
C_s(\alpha,\gamma) \leq \sum_{i=\ell+1}^{k}
\min\{(d-i+1)\beta,\alpha\},
\end{equation}
\end{theorem}
where $\gamma = d\beta$.

Next, we consider an important operational regime, namely the
\emph{bandwidth-limited regime}, where the repair bandwidth $\gamma$
is constrained to a maximum amount $\Gamma$, i.e., $\gamma \leq
\Gamma$, while no constraint is imposed on the storage capacity
$\alpha$ at each node. The secrecy capacity in this regime is defined
as,
\begin{equation}
C_s^{BL}(\Gamma):=\sup_{\begin{array}{c}
 \gamma \leq \Gamma, 0 \leq \alpha
\end{array}
} C_s(\alpha,\gamma).
\end{equation}
For a fixed $\Gamma$, when the parameter $d$ is a system design
choice, the upper bound of Theorem~\ref{thm:converse} on the secrecy
capacity can be further optimized, and attains a maximum for
$d=n-1$. In section~\ref{sec:achievability}, we demonstrate that this
upper bound  can be achieved for $d = n-1$ in the bandwidth-limited
regime. Thus, establishing the following theorem:
\begin{theorem}\label{thm:BW_limited}[Bandwidth-Limited Regime]
For a distributed data storage system $\mathcal{D}(n,k)$,
$\ell<k$ compromised nodes, the secrecy capacity for a
bandwidth-limited regime, for $d=n-1$, is
\begin{equation}
C_s^{BL}(\Gamma) = \sum_{i=\ell+1}^{k}(n-i)\frac{\Gamma}{n-1},
\end{equation}
and is achieved with a storage capacity of $\alpha= \Gamma$.
\end{theorem}


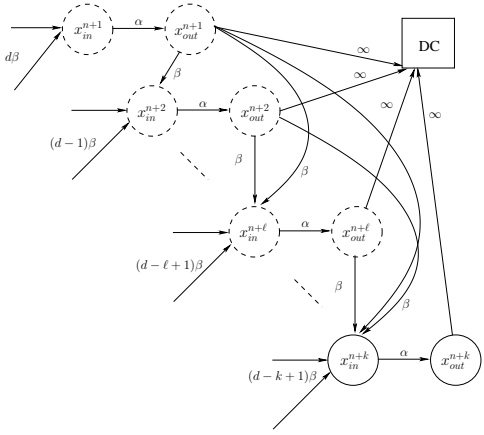
\begin{figure}[t]
\begin{center}
\resizebox{2.5in}{!} {\input{passive_ub.pstex_t}} \caption{Part of
the flow graph corresponding to a DSS ${\cal D}(n,k)$, when nodes
$v_1,\dots,v_k$ fail successively, and are replaced by nodes
$v_{n+1},\dots,v_{n+k}$. A data collector DC connects to these $k$ nodes and
wants to retrieve the whole file. Nodes $v_{n+1},\dots,v_{n+\ell}$
shown with broken boundaries are compromised by Eve during
repair.}\label{fig:converse}\vsp
\end{center}
\end{figure}

\section{Special Cases}\label{sec:special}%
\subsection{Static Systems}
A static version of the problem studied here corresponds to a DSS
with ideal storage nodes that do not fail. Hence there is no need
for any repair in the system. The flow graph of this system is then
the combination network studied in network coding theory (see for
e.g. \cite[Chap. 4]{YLC06} ). Therefore, the static storage problem
can be regarded as a special case of wiretap networks \cite{CY02,
RS07}, or equivalently, as the erasure-erasure wiretap-II channel
studied in \cite{AM09}. The secrecy capacity for such systems is
$(k-\ell)\alpha$, and can be achieved using either nested MDS codes
 \cite{AM09}, or the coset codes of \cite{OW,RS07}.

Even though the above proposed solution is optimal for the static
case, it can have a very poor secrecy performance when applied
directly to dynamic storage systems with failures. For instance, a
straightforward way to repair a failed node would be to download the
whole file on the new replacement node, and then generate the
specific lost data. In this case, if Eve accesses the new replacement
node while it is downloading the whole file, it will be able to
reconstruct the entire original data. Hence, the secrecy rate for
this scheme would be zero. However, Theorem~\ref{thm:BW_limited}
suggests that for some systems we can achieve a positive secrecy
capacity. This example highlights the fact that dynamical repair of
the DSS renders it intrinsically different from the static
counterpart, and one should be careful in designing the repair scheme
in order to safeguard the whole stored data.

\subsection{Systems Using Random Network Coding}\label{sec:RNC}
Using the flow graph model, the authors of \cite{DGWWR07} showed that
{\em random linear network codes} over a large finite field can
achieve any point $(\alpha,\gamma)$, on the optimal storage-repair
bandwidth tradeoff curve with a high probability. Consider an example
of random linear network code used in a compromised DSS
$\mathcal{D}(4,3)$, which stores $R=6$ symbols and operates at
$d=3,\beta = 1,$ and $\alpha=3$. In this case, each of the initial
nodes $v_1,\dots,v_4$ stores $3$ independently generated random
linear combinations of these $R=6$ symbols. Assume now that node
$v_4$ fails and is replaced by a new node $v_5$ that connects to
$v_1,v_2$, and $v_3$, and downloads from each one of them $\beta=1$
random linear combination of their stored data. Assume that after
some time, node $v_5$ fails and is replaced by node $v_6$ in a
similar fashion. Now, if $\ell = 2$, and Eve accesses nodes $v_5$ and
$v_6$ while they were being repaired, it will observe 6 linear
combinations of the original data symbols, which, with high
probability are linearly independent. Therefore, she will be able to
reconstruct the whole file.

The above analysis shows that, when random network coding is used, it
is not possible to achieve a positive secrecy rate for this system,
even with pre-processing at the source, using for example Maximum Rank
Distance (MRD) codes \cite{SK08}. But according to
Theorem~\ref{thm:BW_limited}, which we prove in
section~\ref{sec:achievability}, the secrecy capacity of the the
above DSS $D(4,3)$ is equal to one unit when $\ell=2$. This is also
in contrast with the case of multicast networks with compromised
edges instead of nodes \cite{CY02}, wherein, random network coding
can perform as good as any deterministic secure code \cite{SK08}.

\section{Upper bound on secrecy capacity}\label{sec:UB_capacity}
In this section we derive the upper bound of
Theorem~\ref{thm:converse}. Consider a DSS $\mathcal{D}(n,k)$ with
$\ell < k$. Assume that the nodes $v_1, v_2,\dots,v_k$ have failed
consecutively, and were replaced during the repair process by the
nodes $v_{n+1},v_{n+2},\dots,v_{n+k}$ respectively as shown in
Fig.~\ref{fig:converse}. Now suppose that Eve accesses nodes in
$E=\{v_{n+1},v_{n+2},\hdots,v_{n+\ell}\}$ while they were being
repaired, and consider a data collector connected to the nodes in
$B=\{v_{n+1},v_{n+2},\hdots,v_{n+k}\}$. The reconstruction property
implies $H(S|C_B)=0$ by Eq.~\eqref{eq:reconstruction_condition}, and
the perfect secrecy condition implies $H(S|D_E)=H(S)$ by
Eq.~\eqref{eq:secrecy_condition}. We can therefore write
\begin{equation*}
\begin{split}
H(S) & = H(S|D_E)-H(S|C_B)\\
& \overset{(1)}\leq H(S|C_E)-H(S|C_B)\\
& \overset{(2)}= H(S|C_E) - H(S|C_E,C_{B\setminus E})\\
&= I(S,C_{B\setminus E}|C_E)\\
&\leq H(C_{B\setminus E}|C_E)\\
&= \sum_{i=\ell+1}^k H(C_{n+i}|C_{n+1},\dots,C_{n+i-1})\\
&\overset{(3)}\leq \sum_{i=\ell+1}^{k} \min\{(d-i+1)\beta,\alpha\}.
\end{split}
\end{equation*}
Inequality $(1)$ follows from the fact that the stored data $C_E$ is
a function of the downloaded data $D_E$, (2) from, $C_{B\setminus
E}:=\{C_{n+\ell+1},\dots,C_{n+k}\}$, (3) follows from the fact that
each node can store at most $\alpha$ units, and for each replacement
node we have $H(C_i) \leq H(D_i) \leq d\beta$, also from the topology
of the network (see Fig.~\ref{fig:converse}). Note that each node
$x_{in}^{n+i}$ is connected to each of the nodes
$x_{out}^{n+1},\dots,x_{out}^{n+i-1}$ by an edge of capacity $\beta$.
The upper bound of Theorem~\ref{thm:converse} follows then directly
from the definition of Eq.~\eqref{eq:secrecy_capacity}.

\section{Secrecy Capacity in the Bandwidth-Limited Regime}\label{sec:achievability}
\subsection{Example}\label{sec:DSS_passive_ex}
\begin{figure}[t]
\begin{center}
\resizebox{2.8in}{!} {\input{dss_coset_ex.pstex_t}}
\caption{Schematic representation of the optimal code for the DSS ${\cal
D}(4,3)$ with $\alpha=3,\beta=1,d=3$, and $\ell =2$ that achieves the
secrecy capacity of $1$ unit. An MDS coset code takes the information symbol $S$
and five independent random keys $K_1,\dots,K_5$, as an input and outputs a parity check symbol $Z=S+\sum_{i=1}^5K_i$, along with random keys in systematic form.
These symbols are then stored on the DSS using the code
structure of \cite{RSKR09}.}\label{fig:dss_coset_ex}\vsp
\end{center}
\end{figure}
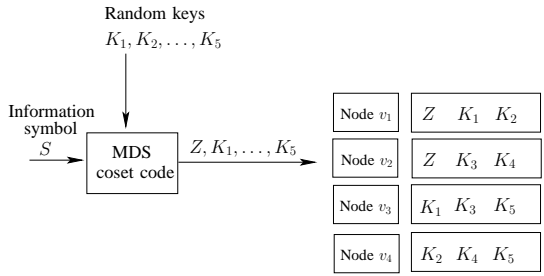

Consider again the DSS $\mathcal{D}(4,3)$ with $\alpha=3,d=3,\beta=1$,
and $\ell =2$ of Section~\ref{sec:RNC}, for which the secrecy rate
using random linear network coding was shown to be $0$. The upper
bound on the secrecy capacity of this system given by
Theorem~\ref{thm:converse} is $1$. We provide a scheme that achieves
this upper bound. The proposed code is depicted in
Fig.~\ref{fig:dss_coset_ex} and consists of the concatenation of an
MDS coset code \cite{OW} with a special repetition code that was
introduced in \cite{RSKR09} by Rashmi et al.\ for constructing exact regeneration
codes. Let $S\in \mathbb{F}_q$ denote the information symbol to be
securely stored on the system. $S$ is encoded using the outer MDS
code into a codeword $(Z,K_1,K_2,\dots,K_5)$, where $K_1,\dots,K_5$
are independent random keys uniformly distributed over $\mathbb{F}_q$
and $Z=S + \sum_{i=1}^{5} K_i$. The encoded symbols $Z,K_1,\dots,K_5$
are then stored on the nodes $v_1,\dots,v_4$ as shown in
Fig.~\ref{fig:dss_coset_ex}, following the special repetition code of
 \cite{RSKR09}. It is easy to verify that any data
collector connecting to $3$ nodes, observes all the symbols
$Z,K_1,\dots, K_5$, and can therefore decode $S=Z-\sum_{i=1}^5 K_i$.
However, an eavesdropper accessing any two nodes will only observe
$5$ symbols out of $6$, and cannot gain any information
about $S$. Next, we generalize this construction to obtain a
capacity-achieving code for the bandwidth-limited regime.

\subsection{Code Construction}\label{sec:passive_achievability}
Our approach builds on the results of \cite{RSKR09} where the authors
constructed a family of exact regenerating codes for $d=n-1$. The
``exact" property of these codes allows any repair node to
reconstruct and store an identical copy of the data lost upon a
failure. For simplicity, we will explain the construction for $\beta
= 1$, i.e., $\Gamma=n-1$. For any larger values of $\Gamma$, and in
turn of $\beta$, the file can be split into chunks, each of which can
be separately encoded using the construction corresponding to $\beta
= 1$. Choose $\alpha = \Gamma$. From \cite{DGWWR07} we know that $M=
\sum_{i=1}^{k} (n-i)$ is the capacity of the above DSS in the absence
of any adversary ($\ell=0$). Let $R:=\sum_{i=\ell+1}^{k} (n-i)$ be
the number of information symbols that we would like to store
securely on the DSS, and $\theta:=\frac{n(n-1)}{2}$. Let
$S=(s_1,\dots,s_R)\in \mathbb{F}_q^R$ denote the information file and
${\cal K} = (K_1,\dots,K_{M-R}) \in \mathbb{F}_q^{M-R}$ denote $M-R$
independent random keys each uniformly distributed over
$\mathbb{F}_q$. Then, the proposed code consists of an outer nested
$(\theta,M)$ MDS coset code \cite{AM09} which takes $S$ and ${\cal K}$ as
an input, and outputs $X=(x_1,\dots,x_\theta )$, such that $X = {\cal K} G_K
+ S G_S$, where $G = \left(
       \begin{array}{c}
        G_K \\
        G_S \\
       \end{array}
      \right)
$ is a generator matrix of a $(\theta,M)$ MDS code, and $G_K$ in
itself is a generator matrix for a $(\theta,M-R)$ MDS code. The
information vector $S$ effectively selects the coset of the MDS code
generated by $G_K$.

This outer $(\theta,M)$ MDS code is then followed by the special
repetition code introduced in \cite{RSKR09} which stores the codeword
$X$ on the DSS. The procedure of constructing this inner code can be
 described using an auxiliary complete graph over $n$
vertices $u_1,\dots,u_n$ that consists of $\theta $ edges. Suppose
the edges are indexed by the coded symbols $x_1,\dots,x_\theta$. The
code then consists of storing on node $v_i$ the indices of the edges
adjacent to vertex $u_i$ in the complete graph. Consequently, every
coded symbol $x_i$ is stored on exactly two storage nodes, and any
pair of two storage nodes have exactly one distinct coded symbol in
common, e.g., code in Fig.~\ref{fig:dss_coset_ex} for $n=4$.

This inner code transforms the dynamic storage system into an equivalent static
point-to-point channel. First notice that $\alpha = \Gamma$, hence
all the data downloaded during the repair process, i.e., $d\beta =
\Gamma$, is stored on the new replacement node without any further
compression. Thus, accessing a node during repair process, i.e.,
observing its downloaded data, is equivalent to accessing it after
the repair process, i.e., observing its stored data. Second, the
exact regeneration codes restore a failed node with the exact lost
data. So, even though there are failures and repairs, the data storage
system looks exactly the same at any point of time. Any data
collector downloads $M$ symbols out of $x_1,\dots,x_\theta$ by
connecting to $k$ nodes. Moreover, any eavesdropper can observe
$\mu=\sum_{i=1}^{\ell} (n-i) = M - R$ symbols. Thus, the system
becomes similar to the erasure-erasure wiretap channel-II of
parameters $(\theta,M,\mu)$\footnote{ In the erasure-erasure
wiretap channel-II of parameters $(\theta,M,\mu)$, the transmitter
sends $\theta$ symbols. A legitimate receiver and an eavesdropper
receive $M$ and $\mu$ symbols respectively through independent erasure
channels \cite{AM09}.}. Therefore, since the outer code is a nested
MDS code, from \cite{AM09} we know that it can achieve the secrecy
capacity of $M-\mu=M - (M-R) = R = \sum_{i=\ell+1}^{k} (n-i)$ of the
corresponding erasure-erasure wiretap channel. This rate is achieved
for every $1$ unit of $\beta$. Thus, the total secrecy rate achieved
for $\beta = \Gamma/(n-1)$ is $\sum_{i=\ell+1}^{k}
(n-i)\frac{\Gamma}{n-1}$.

\section{Conclusion}\label{sec:con}
In this paper we considered dynamic distributed data storage systems
that are subject to eavesdropping. Our main objective was to
determine the secrecy capacity of such systems, i.e., the maximum
amount of data that these systems can store and deliver to data
collectors, without revealing any information to the eavesdropper.
Modeling such systems as multicast networks with compromised nodes,
we gave an upper bound on the secrecy capacity, and showed that it can
be achieved in the important bandwidth-limited regime
where the nodes have sufficient storage capacity. Finding the
general expression of the secrecy capacity of distributed storage
systems, and more generally of multicast networks with a fixed number
of compromised nodes, remains an open problem that we hope to address
in future work.
\bibliographystyle{ieeetr}
\bibliography{DSS}
\end{document}

%% file: dss_example.pstex_t
\begin{picture}(0,0)%
\includegraphics{dss_example.pstex}%
\end{picture}%
\setlength{\unitlength}{3947sp}%
\begingroup\makeatletter\ifx\SetFigFont\undefined%
\gdef\SetFigFont#1#2#3#4#5{%
  \reset@font\fontsize{#1}{#2pt}%
  \fontfamily{#3}\fontseries{#4}\fontshape{#5}%
  \selectfont}%
\fi\endgroup%
\begin{picture}(7077,7539)(171,-8114)
\put(3012,-7357){\makebox(0,0)[lb]{\smash{{\SetFigFont{17}{20.4}{\rmdefault}{\mddefault}{\updefault}{\color[rgb]{0,0,0}$a_1+2b_1$}%
}}}}
\put(2975,-7707){\makebox(0,0)[lb]{\smash{{\SetFigFont{17}{20.4}{\rmdefault}{\mddefault}{\updefault}{\color[rgb]{0,0,0}$2a_2+b_2$}%
}}}}
\put(3277,-6682){\makebox(0,0)[lb]{\smash{{\SetFigFont{17}{20.4}{\rmdefault}{\mddefault}{\updefault}{\color[rgb]{0,0,0}$v_4$}%
}}}}
\put(5838,-3397){\makebox(0,0)[lb]{\smash{{\SetFigFont{17}{20.4}{\rmdefault}{\mddefault}{\updefault}{\color[rgb]{0,0,0}$a_1+a_2$}%
}}}}
\put(5813,-3734){\makebox(0,0)[lb]{\smash{{\SetFigFont{17}{20.4}{\rmdefault}{\mddefault}{\updefault}{\color[rgb]{0,0,0}$a_1+4a_2$}%
}}}}
\put(6176,-2697){\makebox(0,0)[lb]{\smash{{\SetFigFont{17}{20.4}{\rmdefault}{\mddefault}{\updefault}{\color[rgb]{0,0,0}$v_5$}%
}}}}
\put(6526,-6290){\makebox(0,0)[lb]{\smash{{\SetFigFont{17}{20.4}{\rmdefault}{\mddefault}{\updefault}{\color[rgb]{0,0,0}DC}%
}}}}
\put(3276,-4705){\makebox(0,0)[lb]{\smash{{\SetFigFont{17}{20.4}{\rmdefault}{\mddefault}{\updefault}{\color[rgb]{0,0,0}$v_3$}%
}}}}
\put(3013,-5730){\makebox(0,0)[lb]{\smash{{\SetFigFont{17}{20.4}{\rmdefault}{\mddefault}{\updefault}{\color[rgb]{0,0,0}$a_2+b_2$}%
}}}}
\put(3013,-5417){\makebox(0,0)[lb]{\smash{{\SetFigFont{17}{20.4}{\rmdefault}{\mddefault}{\updefault}{\color[rgb]{0,0,0}$a_1+b_1$}%
}}}}
\put(364,-3411){\makebox(0,0)[lb]{\smash{{\SetFigFont{17}{20.4}{\rmdefault}{\mddefault}{\updefault}{\color[rgb]{0,0,0}File ${\cal F}$}%
}}}}
\put(3302,-841){\makebox(0,0)[lb]{\smash{{\SetFigFont{17}{20.4}{\rmdefault}{\mddefault}{\updefault}{\color[rgb]{0,0,0}$v_1$}%
}}}}
\put(3314,-2733){\makebox(0,0)[lb]{\smash{{\SetFigFont{17}{20.4}{\rmdefault}{\mddefault}{\updefault}{\color[rgb]{0,0,0}$v_2$}%
}}}}
\put(3226,-1686){\makebox(0,0)[lb]{\smash{{\SetFigFont{17}{20.4}{\rmdefault}{\mddefault}{\updefault}{\color[rgb]{0,0,0}$a_1, a_2$}%
}}}}
\put(3214,-3599){\makebox(0,0)[lb]{\smash{{\SetFigFont{17}{20.4}{\rmdefault}{\mddefault}{\updefault}{\color[rgb]{0,0,0}$b_1, b_2$}%
}}}}
\put(526,-4561){\makebox(0,0)[lb]{\smash{{\SetFigFont{17}{20.4}{\rmdefault}{\mddefault}{\updefault}{\color[rgb]{0,0,0}$b_1, b_2$}%
}}}}
\put(514,-4186){\makebox(0,0)[lb]{\smash{{\SetFigFont{17}{20.4}{\rmdefault}{\mddefault}{\updefault}{\color[rgb]{0,0,0}$a_1, a_2$}%
}}}}
\put(4976,-3386){\makebox(0,0)[lb]{\smash{{\SetFigFont{17}{20.4}{\rmdefault}{\mddefault}{\updefault}{\color[rgb]{0,0,0}1}%
}}}}
\put(2026,-2824){\makebox(0,0)[lb]{\smash{{\SetFigFont{17}{20.4}{\rmdefault}{\mddefault}{\updefault}{\color[rgb]{0,0,0}2}%
}}}}
\put(2176,-3774){\makebox(0,0)[lb]{\smash{{\SetFigFont{17}{20.4}{\rmdefault}{\mddefault}{\updefault}{\color[rgb]{0,0,0}2}%
}}}}
\put(2264,-4761){\makebox(0,0)[lb]{\smash{{\SetFigFont{17}{20.4}{\rmdefault}{\mddefault}{\updefault}{\color[rgb]{0,0,0}2}%
}}}}
\put(2214,-5724){\makebox(0,0)[lb]{\smash{{\SetFigFont{17}{20.4}{\rmdefault}{\mddefault}{\updefault}{\color[rgb]{0,0,0}2}%
}}}}
\put(5176,-4099){\makebox(0,0)[lb]{\smash{{\SetFigFont{17}{20.4}{\rmdefault}{\mddefault}{\updefault}{\color[rgb]{0,0,0}1}%
}}}}
\put(5601,-4649){\makebox(0,0)[lb]{\smash{{\SetFigFont{17}{20.4}{\rmdefault}{\mddefault}{\updefault}{\color[rgb]{0,0,0}1}%
}}}}
\put(5489,-6911){\makebox(0,0)[lb]{\smash{{\SetFigFont{17}{20.4}{\rmdefault}{\mddefault}{\updefault}{\color[rgb]{0,0,0}$\infty$}%
}}}}
\put(5714,-5311){\makebox(0,0)[lb]{\smash{{\SetFigFont{17}{20.4}{\rmdefault}{\mddefault}{\updefault}{\color[rgb]{0,0,0}$\infty$}%
}}}}
\end{picture}%

%% file: flowgraph_ex.pstex_t
\begin{picture}(0,0)%
\includegraphics{flowgraph_ex.pstex}%
\end{picture}%
\setlength{\unitlength}{3947sp}%
\begingroup\makeatletter\ifx\SetFigFont\undefined%
\gdef\SetFigFont#1#2#3#4#5{%
  \reset@font\fontsize{#1}{#2pt}%
  \fontfamily{#3}\fontseries{#4}\fontshape{#5}%
  \selectfont}%
\fi\endgroup%
\begin{picture}(6853,6237)(1458,-6863)
\put(3850,-4482){\makebox(0,0)[lb]{\smash{{\SetFigFont{14}{16.8}{\rmdefault}{\mddefault}{\updefault}{\color[rgb]{0,0,0}$\alpha$}%
}}}}
\put(3837,-6144){\makebox(0,0)[lb]{\smash{{\SetFigFont{14}{16.8}{\rmdefault}{\mddefault}{\updefault}{\color[rgb]{0,0,0}$\alpha$}%
}}}}
\put(2276,-2397){\makebox(0,0)[lb]{\smash{{\SetFigFont{14}{16.8}{\rmdefault}{\mddefault}{\updefault}{\color[rgb]{0,0,0}$\infty$}%
}}}}
\put(2588,-3847){\makebox(0,0)[lb]{\smash{{\SetFigFont{14}{16.8}{\rmdefault}{\mddefault}{\updefault}{\color[rgb]{0,0,0}$\infty$}%
}}}}
\put(2213,-4910){\makebox(0,0)[lb]{\smash{{\SetFigFont{14}{16.8}{\rmdefault}{\mddefault}{\updefault}{\color[rgb]{0,0,0}$\infty$}%
}}}}
\put(2501,-2947){\makebox(0,0)[lb]{\smash{{\SetFigFont{14}{16.8}{\rmdefault}{\mddefault}{\updefault}{\color[rgb]{0,0,0}$\infty$}%
}}}}
\put(3838,-5060){\makebox(0,0)[lb]{\smash{{\SetFigFont{20}{24.0}{\rmdefault}{\mddefault}{\updefault}{\color[rgb]{0,0,0}$v_3$}%
}}}}
\put(3851,-3435){\makebox(0,0)[lb]{\smash{{\SetFigFont{20}{24.0}{\rmdefault}{\mddefault}{\updefault}{\color[rgb]{0,0,0}$v_2$}%
}}}}
\put(3838,-6747){\makebox(0,0)[lb]{\smash{{\SetFigFont{20}{24.0}{\rmdefault}{\mddefault}{\updefault}{\color[rgb]{0,0,0}$v_4$}%
}}}}
\put(3863,-1169){\makebox(0,0)[lb]{\smash{{\SetFigFont{14}{16.8}{\rmdefault}{\mddefault}{\updefault}{\color[rgb]{0,0,0}$\alpha$}%
}}}}
\put(3850,-2832){\makebox(0,0)[lb]{\smash{{\SetFigFont{14}{16.8}{\rmdefault}{\mddefault}{\updefault}{\color[rgb]{0,0,0}$\alpha$}%
}}}}
\put(7150,-2819){\makebox(0,0)[lb]{\smash{{\SetFigFont{14}{16.8}{\rmdefault}{\mddefault}{\updefault}{\color[rgb]{0,0,0}$\alpha$}%
}}}}
\put(6139,-3786){\makebox(0,0)[lb]{\smash{{\SetFigFont{14}{16.8}{\rmdefault}{\mddefault}{\updefault}{\color[rgb]{0,0,0}$\beta = 1$}%
}}}}
\put(5676,-3411){\makebox(0,0)[lb]{\smash{{\SetFigFont{14}{16.8}{\rmdefault}{\mddefault}{\updefault}{\color[rgb]{0,0,0}$\beta$}%
}}}}
\put(5551,-2836){\makebox(0,0)[lb]{\smash{{\SetFigFont{14}{16.8}{\rmdefault}{\mddefault}{\updefault}{\color[rgb]{0,0,0}$\beta = 1$}%
}}}}
\put(3863,-1810){\makebox(0,0)[lb]{\smash{{\SetFigFont{20}{24.0}{\rmdefault}{\mddefault}{\updefault}{\color[rgb]{0,0,0}$v_1$}%
}}}}
\put(7126,-3399){\makebox(0,0)[lb]{\smash{{\SetFigFont{20}{24.0}{\rmdefault}{\mddefault}{\updefault}{\color[rgb]{0,0,0}$v_5$}%
}}}}
\put(6451,-5799){\makebox(0,0)[lb]{\smash{{\SetFigFont{14}{16.8}{\rmdefault}{\mddefault}{\updefault}{\color[rgb]{0,0,0}$\infty$}%
}}}}
\put(6564,-4399){\makebox(0,0)[lb]{\smash{{\SetFigFont{14}{16.8}{\rmdefault}{\mddefault}{\updefault}{\color[rgb]{0,0,0}$\infty$}%
}}}}
\put(7489,-5286){\makebox(0,0)[lb]{\smash{{\SetFigFont{17}{20.4}{\rmdefault}{\mddefault}{\updefault}{\color[rgb]{0,0,0}DC}%
}}}}
\put(3139,-1374){\makebox(0,0)[lb]{\smash{{\SetFigFont{17}{20.4}{\rmdefault}{\mddefault}{\updefault}{\color[rgb]{0,0,0}$x_{in}^1$}%
}}}}
\put(3126,-4686){\makebox(0,0)[lb]{\smash{{\SetFigFont{17}{20.4}{\rmdefault}{\mddefault}{\updefault}{\color[rgb]{0,0,0}$x_{in}^3$}%
}}}}
\put(6426,-3036){\makebox(0,0)[lb]{\smash{{\SetFigFont{17}{20.4}{\rmdefault}{\mddefault}{\updefault}{\color[rgb]{0,0,0}$x_{in}^5$}%
}}}}
\put(3114,-6361){\makebox(0,0)[lb]{\smash{{\SetFigFont{17}{20.4}{\rmdefault}{\mddefault}{\updefault}{\color[rgb]{0,0,0}$x_{in}^4$}%
}}}}
\put(3126,-3049){\makebox(0,0)[lb]{\smash{{\SetFigFont{17}{20.4}{\rmdefault}{\mddefault}{\updefault}{\color[rgb]{0,0,0}$x_{in}^2$}%
}}}}
\put(1764,-3336){\makebox(0,0)[lb]{\smash{{\SetFigFont{17}{20.4}{\rmdefault}{\mddefault}{\updefault}{\color[rgb]{0,0,0}$s$}%
}}}}
\put(4439,-6374){\makebox(0,0)[lb]{\smash{{\SetFigFont{17}{20.4}{\rmdefault}{\mddefault}{\updefault}{\color[rgb]{0,0,0}$x_{out}^4$}%
}}}}
\put(4464,-1374){\makebox(0,0)[lb]{\smash{{\SetFigFont{17}{20.4}{\rmdefault}{\mddefault}{\updefault}{\color[rgb]{0,0,0}$x_{out}^1$}%
}}}}
\put(4439,-4699){\makebox(0,0)[lb]{\smash{{\SetFigFont{17}{20.4}{\rmdefault}{\mddefault}{\updefault}{\color[rgb]{0,0,0}$x_{out}^3$}%
}}}}
\put(4439,-3061){\makebox(0,0)[lb]{\smash{{\SetFigFont{17}{20.4}{\rmdefault}{\mddefault}{\updefault}{\color[rgb]{0,0,0}$x_{out}^2$}%
}}}}
\put(7714,-3036){\makebox(0,0)[lb]{\smash{{\SetFigFont{17}{20.4}{\rmdefault}{\mddefault}{\updefault}{\color[rgb]{0,0,0}$x_{out}^5$}%
}}}}
\end{picture}%

%% file: passive_ub.pstex_t
\begin{picture}(0,0)%
\includegraphics{passive_ub.pstex}%
\end{picture}%
\setlength{\unitlength}{3947sp}%
\begingroup\makeatletter\ifx\SetFigFont\undefined%
\gdef\SetFigFont#1#2#3#4#5{%
  \reset@font\fontsize{#1}{#2pt}%
  \fontfamily{#3}\fontseries{#4}\fontshape{#5}%
  \selectfont}%
\fi\endgroup%
\begin{picture}(8885,7941)(349,-8523)
\put(364,-1661){\makebox(0,0)[lb]{\smash{{\SetFigFont{14}{16.8}{\rmdefault}{\mddefault}{\updefault}{\color[rgb]{0,0,0}$d\beta$}%
}}}}
\put(2776,-986){\makebox(0,0)[lb]{\smash{{\SetFigFont{14}{16.8}{\rmdefault}{\mddefault}{\updefault}{\color[rgb]{0,0,0}$\alpha$}%
}}}}
\put(5876,-4755){\makebox(0,0)[lb]{\smash{{\SetFigFont{14}{16.8}{\rmdefault}{\mddefault}{\updefault}{\color[rgb]{0,0,0}$\alpha$}%
}}}}
\put(2801,-5536){\makebox(0,0)[lb]{\smash{{\SetFigFont{14}{16.8}{\rmdefault}{\mddefault}{\updefault}{\color[rgb]{0,0,0}$(d-\ell + 1)\beta$}%
}}}}
\put(4889,-7599){\makebox(0,0)[lb]{\smash{{\SetFigFont{14}{16.8}{\rmdefault}{\mddefault}{\updefault}{\color[rgb]{0,0,0}$(d-k+1)\beta$}%
}}}}
\put(1201,-3261){\makebox(0,0)[lb]{\smash{{\SetFigFont{14}{16.8}{\rmdefault}{\mddefault}{\updefault}{\color[rgb]{0,0,0}$(d-1)\beta$}%
}}}}
\put(3976,-2505){\makebox(0,0)[lb]{\smash{{\SetFigFont{14}{16.8}{\rmdefault}{\mddefault}{\updefault}{\color[rgb]{0,0,0}$\alpha$}%
}}}}
\put(7714,-7144){\makebox(0,0)[lb]{\smash{{\SetFigFont{14}{16.8}{\rmdefault}{\mddefault}{\updefault}{\color[rgb]{0,0,0}$\alpha$}%
}}}}
\put(3501,-1949){\makebox(0,0)[lb]{\smash{{\SetFigFont{14}{16.8}{\rmdefault}{\mddefault}{\updefault}{\color[rgb]{0,0,0}$\beta$}%
}}}}
\put(4639,-3561){\makebox(0,0)[lb]{\smash{{\SetFigFont{14}{16.8}{\rmdefault}{\mddefault}{\updefault}{\color[rgb]{0,0,0}$\beta$}%
}}}}
\put(6501,-5911){\makebox(0,0)[lb]{\smash{{\SetFigFont{14}{16.8}{\rmdefault}{\mddefault}{\updefault}{\color[rgb]{0,0,0}$\beta$}%
}}}}
\put(6914,-1499){\makebox(0,0)[lb]{\smash{{\SetFigFont{14}{16.8}{\rmdefault}{\mddefault}{\updefault}{\color[rgb]{0,0,0}$\infty$}%
}}}}
\put(6851,-1974){\makebox(0,0)[lb]{\smash{{\SetFigFont{14}{16.8}{\rmdefault}{\mddefault}{\updefault}{\color[rgb]{0,0,0}$\infty$}%
}}}}
\put(7351,-2536){\makebox(0,0)[lb]{\smash{{\SetFigFont{14}{16.8}{\rmdefault}{\mddefault}{\updefault}{\color[rgb]{0,0,0}$\infty$}%
}}}}
\put(8251,-2711){\makebox(0,0)[lb]{\smash{{\SetFigFont{14}{16.8}{\rmdefault}{\mddefault}{\updefault}{\color[rgb]{0,0,0}$\infty$}%
}}}}
\put(5864,-3786){\makebox(0,0)[lb]{\smash{{\SetFigFont{14}{16.8}{\rmdefault}{\mddefault}{\updefault}{\color[rgb]{0,0,0}$\beta$}%
}}}}
\put(7751,-6261){\makebox(0,0)[lb]{\smash{{\SetFigFont{14}{16.8}{\rmdefault}{\mddefault}{\updefault}{\color[rgb]{0,0,0}$\beta$}%
}}}}
\put(8064,-1461){\makebox(0,0)[lb]{\smash{{\SetFigFont{17}{20.4}{\rmdefault}{\mddefault}{\updefault}{\color[rgb]{0,0,0}DC}%
}}}}
\put(1651,-1161){\makebox(0,0)[lb]{\smash{{\SetFigFont{17}{20.4}{\rmdefault}{\mddefault}{\updefault}{\color[rgb]{0,0,0}$x^{n+1}_{in}$}%
}}}}
\put(2826,-2699){\makebox(0,0)[lb]{\smash{{\SetFigFont{17}{20.4}{\rmdefault}{\mddefault}{\updefault}{\color[rgb]{0,0,0}$x^{n+2}_{in}$}%
}}}}
\put(4726,-4936){\makebox(0,0)[lb]{\smash{{\SetFigFont{17}{20.4}{\rmdefault}{\mddefault}{\updefault}{\color[rgb]{0,0,0}$x^{n+\ell}_{in}$}%
}}}}
\put(6601,-7311){\makebox(0,0)[lb]{\smash{{\SetFigFont{17}{20.4}{\rmdefault}{\mddefault}{\updefault}{\color[rgb]{0,0,0}$x^{n+k}_{in}$}%
}}}}
\put(3514,-1174){\makebox(0,0)[lb]{\smash{{\SetFigFont{17}{20.4}{\rmdefault}{\mddefault}{\updefault}{\color[rgb]{0,0,0}$x^{n+1}_{out}$}%
}}}}
\put(6639,-4936){\makebox(0,0)[lb]{\smash{{\SetFigFont{17}{20.4}{\rmdefault}{\mddefault}{\updefault}{\color[rgb]{0,0,0}$x^{n+\ell}_{out}$}%
}}}}
\put(4751,-2699){\makebox(0,0)[lb]{\smash{{\SetFigFont{17}{20.4}{\rmdefault}{\mddefault}{\updefault}{\color[rgb]{0,0,0}$x^{n+2}_{out}$}%
}}}}
\put(8476,-7324){\makebox(0,0)[lb]{\smash{{\SetFigFont{17}{20.4}{\rmdefault}{\mddefault}{\updefault}{\color[rgb]{0,0,0}$x^{n+k}_{out}$}%
}}}}
\end{picture}%

%% file: dss_coset_ex.pstex_t
\begin{picture}(0,0)%
\includegraphics{dss_coset_ex.pstex}%
\end{picture}%
\setlength{\unitlength}{3947sp}%
\begingroup\makeatletter\ifx\SetFigFont\undefined%
\gdef\SetFigFont#1#2#3#4#5{%
  \reset@font\fontsize{#1}{#2pt}%
  \fontfamily{#3}\fontseries{#4}\fontshape{#5}%
  \selectfont}%
\fi\endgroup%
\begin{picture}(9787,4957)(49,-5008)
\put(1676,-3261){\makebox(0,0)[lb]{\smash{{\SetFigFont{20}{24.0}{\rmdefault}{\mddefault}{\updefault}{\color[rgb]{0,0,0}coset code}%
}}}}
\put(1964,-2886){\makebox(0,0)[lb]{\smash{{\SetFigFont{20}{24.0}{\rmdefault}{\mddefault}{\updefault}{\color[rgb]{0,0,0}MDS}%
}}}}
\put(626,-2824){\makebox(0,0)[lb]{\smash{{\SetFigFont{20}{24.0}{\rmdefault}{\mddefault}{\updefault}{\color[rgb]{0,0,0}$S$}%
}}}}
\put(1826,-812){\makebox(0,0)[lb]{\smash{{\SetFigFont{20}{24.0}{\rmdefault}{\mddefault}{\updefault}{\color[rgb]{0,0,0}$K_1,K_2,\hdots,K_5$}%
}}}}
\put(8964,-2199){\makebox(0,0)[lb]{\smash{{\SetFigFont{20}{24.0}{\rmdefault}{\mddefault}{\updefault}{\color[rgb]{0,0,0}$K_2$}%
}}}}
\put(8276,-2199){\makebox(0,0)[lb]{\smash{{\SetFigFont{20}{24.0}{\rmdefault}{\mddefault}{\updefault}{\color[rgb]{0,0,0}$K_1$}%
}}}}
\put(7589,-3886){\makebox(0,0)[lb]{\smash{{\SetFigFont{20}{24.0}{\rmdefault}{\mddefault}{\updefault}{\color[rgb]{0,0,0}$K_1$}%
}}}}
\put(8214,-3861){\makebox(0,0)[lb]{\smash{{\SetFigFont{20}{24.0}{\rmdefault}{\mddefault}{\updefault}{\color[rgb]{0,0,0}$K_3$}%
}}}}
\put(8939,-3861){\makebox(0,0)[lb]{\smash{{\SetFigFont{20}{24.0}{\rmdefault}{\mddefault}{\updefault}{\color[rgb]{0,0,0}$K_5$}%
}}}}
\put(8914,-4761){\makebox(0,0)[lb]{\smash{{\SetFigFont{20}{24.0}{\rmdefault}{\mddefault}{\updefault}{\color[rgb]{0,0,0}$K_5$}%
}}}}
\put(8251,-4761){\makebox(0,0)[lb]{\smash{{\SetFigFont{20}{24.0}{\rmdefault}{\mddefault}{\updefault}{\color[rgb]{0,0,0}$K_4$}%
}}}}
\put(7601,-4761){\makebox(0,0)[lb]{\smash{{\SetFigFont{20}{24.0}{\rmdefault}{\mddefault}{\updefault}{\color[rgb]{0,0,0}$K_2$}%
}}}}
\put(8239,-3036){\makebox(0,0)[lb]{\smash{{\SetFigFont{20}{24.0}{\rmdefault}{\mddefault}{\updefault}{\color[rgb]{0,0,0}$K_3$}%
}}}}
\put(8939,-3024){\makebox(0,0)[lb]{\smash{{\SetFigFont{20}{24.0}{\rmdefault}{\mddefault}{\updefault}{\color[rgb]{0,0,0}$K_4$}%
}}}}
\put(6151,-3011){\makebox(0,0)[lb]{\smash{{\SetFigFont{17}{20.4}{\rmdefault}{\mddefault}{\updefault}{\color[rgb]{0,0,0}Node $v_2$}%
}}}}
\put(6126,-3861){\makebox(0,0)[lb]{\smash{{\SetFigFont{17}{20.4}{\rmdefault}{\mddefault}{\updefault}{\color[rgb]{0,0,0}Node $v_3$}%
}}}}
\put(6139,-2174){\makebox(0,0)[lb]{\smash{{\SetFigFont{17}{20.4}{\rmdefault}{\mddefault}{\updefault}{\color[rgb]{0,0,0}Node $v_1$}%
}}}}
\put(6139,-4749){\makebox(0,0)[lb]{\smash{{\SetFigFont{17}{20.4}{\rmdefault}{\mddefault}{\updefault}{\color[rgb]{0,0,0}Node $v_4$}%
}}}}
\put(1826,-337){\makebox(0,0)[lb]{\smash{{\SetFigFont{20}{24.0}{\rmdefault}{\mddefault}{\updefault}{\color[rgb]{0,0,0}Random keys}%
}}}}
\put(3376,-2787){\makebox(0,0)[lb]{\smash{{\SetFigFont{20}{24.0}{\rmdefault}{\mddefault}{\updefault}{\color[rgb]{0,0,0}$Z, K_1, \hdots, K_5$}%
}}}}
\put( 64,-2136){\makebox(0,0)[lb]{\smash{{\SetFigFont{20}{24.0}{\rmdefault}{\mddefault}{\updefault}{\color[rgb]{0,0,0}Information}%
}}}}
\put(351,-2449){\makebox(0,0)[lb]{\smash{{\SetFigFont{20}{24.0}{\rmdefault}{\mddefault}{\updefault}{\color[rgb]{0,0,0}symbol}%
}}}}
\put(7626,-2199){\makebox(0,0)[lb]{\smash{{\SetFigFont{20}{24.0}{\rmdefault}{\mddefault}{\updefault}{\color[rgb]{0,0,0}$Z$}%
}}}}
\put(7639,-3036){\makebox(0,0)[lb]{\smash{{\SetFigFont{20}{24.0}{\rmdefault}{\mddefault}{\updefault}{\color[rgb]{0,0,0}$Z$}%
}}}}
\end{picture}%

%% file: OnSecureDistributedDataStorageUnderRepairDynamics.bbl
\begin{thebibliography}{1}

\bibitem{Ocean}
S.~Rhea, C.~Wells, P.~Eaton, D.~Geels, B.~Zhao, H.~Weatherspoon, and
  J.~Kubiatowicz, ``{M}aintenance-free global data storage,'' {\em IEEE
  Internet Computing}, pp.~40--49, 2001.

\bibitem{DGWWR07}
A.~Dimakis, P.~Godfrey, Y.~Wu, M.~Wainright, and K.~Ramchandran, ``Network
  coding for distributed storage systems,'' {\em to appear in IEEE Trans.
  Inform. Theory}.

\bibitem{CY02}
N.~Cai and R.~W. Yeung, ``Secure network coding,'' in {\em Proc. IEEE Int.
  Symp. Inf. Theory (ISIT)}, 2002.

\bibitem{RS07}
S.~{El Rouayheb} and E.~Soljanin, ``On wiretap networks {II},'' in {\em Proc.
  IEEE Int. Symp. Inf. Theory (ISIT)}, (Nice, France), 2007.

\bibitem{SK08}
D.~Silva and F.~Kschischang, ``Security for wiretap networks via rank-metric
  codes,'' in {\em Proc. IEEE Int. Symp. Inf. Theory (ISIT)}, 2008.

\bibitem{YLC06}
R.~Yeung, S.-Y. Li, and N.~Cai, {\em Network Coding Theory (Foundations and
  Trends in Communications and Information Theory)}.
\newblock Now Publishers Inc, 2006.

\bibitem{AM09}
S.~Arunkumar and S.~W. Mclaughlin, ``{M}{D}{S} codes on erasure-erasure
  wire-tap channel,'' in {\em arXiv:0902.3286v1}, 2009.

\bibitem{OW}
L.~H. Ozarow and A.~D. Wyner, ``Wire-tap channel-{II},'' in {\em AT\&T Bell lab
  tech. journal vol. 63, no. 10}, 1984.

\bibitem{RSKR09}
K.~Rashmi, N.~B. Shah, P.~V. Kumar, and K.~Ramchandran, ``Exact regenerating
  codes for distributed storage,'' in {\em Allerton Conference on Control,
  Computing, and Communication, Urbana-Champaign, IL}, 2009.

\end{thebibliography}
